\documentclass{aa}
\usepackage{graphicx}
\usepackage{txfonts}
\begin{document}

\title{A laser-lock concept to reach cm/s-precision in Doppler
  experiments with Fabry-P\'erot wavelength calibrators}

\titlerunning{A Fabry-P\'erot laser-lock concept for cm/s Doppler precision}

   \author{A. Reiners\inst{1}
          \and
          R.K. Banyal\inst{1}
          \and
          R.G. Ulbrich\inst{2}
          }

   \institute{Institut f\"ur Astrophysik, Friedrich-Hund-Platz 1, 37077 G\"ottingen, Germany\\
              \email{Ansgar.Reiners@phys.uni-goettingen.de}
         \and
             IV. Physikalisches Institut, Friedrich-Hund-Platz 1, 37077 G\"ottingen, Germany\\
             }

   \date{Received \dots; accepted \dots}

% \abstract{}{}{}{}{}
% 5 {} token are mandatory

   \abstract{State-of-the-art Doppler experiments require wavelength
     calibration with precision at the cm\,s$^{-1}$ level. A
     low-finesse Fabry-P\'erot interferometer (FPI) can provide a
     wavelength comb with a very large bandwidth as required for
     astronomical experiments, but unavoidable spectral drifts are
     difficult to control. Instead of actively controlling the FPI
     cavity, we propose to passively stabilize the interferometer and
     track the time-dependent cavity length drift externally using the
     $^{87}$Rb\,$D_2$ atomic line. A dual-finesse cavity allows drift
     tracking during observation. In the low-finesse spectral range,
     the cavity provides a comb transmission spectrum tailored to the
     astronomical spectrograph.  The drift of the cavity length is
     monitored in the high-finesse range relative to an external
     standard: a single narrow transmission peak is locked to an
     external cavity diode laser and compared to an atomic frequency
     from a Doppler-free transition. Following standard locking
     schemes, tracking at sub-mm\,s$^{-1}$ precision can be achieved.
     This is several orders of magnitude better than currently planned
     high-precision Doppler experiments, and it allows freedom for
     relaxed designs including the use of a single-finesse
     interferometer under certain conditions.  All components for the
     proposed setup are readily available, rendering this approach
     particularly interesting for upcoming Doppler experiments. We
     also show that the large number of interference modes used in an
     astronomical FPI allows us to unambiguously identify the
     interference mode of each FPI transmission peak defining its
     \emph{absolute} wavelength solution. The accuracy reached in each
     resonance with the laser concept is then defined by the cavity
     length that is determined from the one locked peak and by the
     group velocity dispersion. The latter can vary by several
     100\,m\,s$^{-1}$ over the relevant frequency range and severely
     limits the accuracy of individual peak locations, although their
     interference modes are known. A potential way to determine the
     absolute peak positions is to externally measure the frequency of
     each individual peak with a laser frequency comb (LFC). Thus, the
     concept of laser-locked FPIs may be useful for applying the
     absolute accuracy of an LFC to astronomical spectrographs without
     the need for an LFC at the observatory.}


   \maketitle
%
%________________________________________________________________

\section{Introduction}

High-precision radial velocity (RV) measurements have become a well
established technique for detecting and characterizing planets around
other stars, and they are in the focus for precision experiments like
determing fundamental constants, measuring the cosmic microwave
background temperature, and directly observating the expansion of the
Universe\citep{2013arXiv1310.3163M}.  While the last experiments are
among the main science cases for high-resolution spectroscopy at
telescopes beyond the 10\,m range, the search for exoplanets has
become a focus of many observatories with telescopes on the 4\,m
scale. Instruments at these observatories require a cost-effective
solution for high-precision wavelength calibration.

The gravitational pull exerted by an extrasolar planet causes reflex
motion of the parent star detectable in the form of a small Doppler
shift in the stellar spectra \citep[e.g.,][]{1952Obs....72..199S,
  1998ARA&A..36...57M, 2008PhST..130a4010M}. The major challenge for
the RV method is to carry out high-precision measurements from tiny
Doppler shifts of the host target over long periods of observations:
an RV precision of $\sim10$~m\,s$^{-1}$ can be sufficient to discover
giant planets up to Jupiter or Neptune mass. Detection of low-mass
planets like Earth requires measurements at a precision of
10~cm\,s$^{-1}$ on timescales of years.

The key instrumental factors limiting the RV precision of
spectrographic measurements are uncertainties in the wavelength
calibration and instrument stability, particularly on timescales of
years or decades. Additional uncertainties caused by the astronomical
targets, like stellar activity jitter, are not considered
here. Limiting factors include consequences from the insufficient
stability of the instrument's line spread function owing to varying
illumination (e.g., fiber scrambling, modal noise) and suboptimal
knowledge about the wavelength solution. Traditional ways of
calibrating the spectrograph rely on using fundamental atomic or
molecular reference lines in optical and near infrared regions where
astronomical spectroscopy is carried out. The most commonly used
reference sources include Th- and U-based hollow cathode lamps (HCL),
iodine or other gas absorption cells, and telluric lines in the
Earth's atmosphere \citep[e.g.,][]{mar92,ost96,ker07,red12}. Despite
their simplicity, these calibration sources have certain drawbacks,
such as uneven line spacing, a narrow spectral range, a large
intensity difference between the bright and faint lines, and line
blending and aging effects. All these factors contribute to the final
uncertainty in the wavelength solution \citep{pep08}.

The perfect calibrator for astronomical RV experiments should provide
a dense grid of evenly distributed spectral lines of uniform intensity
in order to maximize the Doppler information content over the full
wavelength range. Today, the ultimate solution for the wavelength
calibration problem is expected to come from laser frequency combs
(LFC) generated by femtosecond pulsed lasers \citep[e.g.,][]{mur07,
  wil10, ste08, 2008EPJD...48...57B, phi12}. The self-referencing of
the laser-comb lines produces exceptional stability and frequency
precision (better than $10^{-15}$) and is referenced to an atomic
standard for very high accuracy. A drawback of most currently
available systems is that the comb lines are very densely packed with
frequency differences of only a few 100~MHz. The astronomical
spectrographs used for RV experiments usually have a resolving power
of 10--100\,GHz, which is insufficient to resolve the narrowly spaced
comb structure. To increase the comb line spacing, a common solution
is to filter out most of the lines with external Fabry-P\'erot (FP)
cavities \citep{ste09, 2008EPJD...48...57B, qui10}. The filtering
process itself requires very high accuracy and can induce noise in the
piezoelectrically controlled cavities. Therefore, exactly fine-tuning
the single-finesse cavity's free spectral range (FSR) to the desired
repetition frequency of the comb lines is demanding. Other technical
challenges include the broadband coating dispersion of the cavity
mirror, causing some mismatch between the comb lines and the cavity
resonance and imperfect filtering arising from a relatively broad
cavity linewidth. The light leakage from the partly attenuated modes
gets amplified in the final stage of the second harmonic generation,
thus shifting the line center observed by the spectrograph
\citep{sch08}. Such technical difficulties are likely to be overcome
during the next years, but the question when LFC technology becomes a
turn-key and affordable solution for astronomical observatories is
still open.

A simple and elegant approach to generating a ``perfect'' reference
spectrum is to use the cavity resonance lines of an FPI illuminated
with white light \citep[and without stabilizing it with an
LFC;][]{wil09,wil10-2,hal12}. A passively stabilized ``ideal'' FPI
that is illuminated with a broadband light source produces
transmission lines that are equidistant in frequency space. The
position, linewidth, spacing, and amplitude of these synthetic lines
can be easily tailored to match the spectrograph requirements. The
challenge for the passive FPI is that frequency drifts can occur for
many reasons, including changing environmental conditions, and that
active control of an FPI tailored for astronomical research is very
difficult at m\,s$^{-1}$ accuracy. A simple solution for astronomical
programs is to cross-check the calibration spectrum of the FPI against
classical calibration methods like HCLs, which can be done during
daytime. This strategy, however, is again limited by the accuracy of
the HCLs (and their practical problems).

Our solution is to passively stabilize the FPI and track the drift on
an absolute scale, for example, relative to an atomic transition. To
reach the high precision and long-term stability of the FPI cavity, we
propose a laser-lock concept to track the dimensional stability of a
dual-finesse FPI cavity using frequency stabilized diode lasers. The
mechanism exploits the high accuracy of optical frequency measurements
and connects it to wavelength standards useful in astronomy. Our
strategy eliminates the need for comparing the FPI signal to gas
emission lines on the CCD and provides absolute calibration during
observation. Such a system has all advantages of the optical frequency
comb but can be built entirely from simpler (and cheaper)
off-the-shelf technology. In this paper, we introduce a concept that
utilizes frequency metrology techniques to track the stability of FP
calibrators for astronomical use. The level of implementation details
is kept to a minimum in order to introduce the concept clearly and to
avoid confusion between technical shortcomings and general
limitations, the latter being the main obstacle in current solutions
for astronomical wavelength calibration.

In Section\,2, we discuss instrument stability and design
requirements. The laser-lock concept and the dual-channel FPI are
described in Sections\,3 and 4, respectively. A discussion of laser
frequency control relevant to laser-locking is provided in
Section\,5. A method for establishing an absolute frequency scale for
FPI calibration is proposed in Section\,6.  Finally, we present a
summary in Section~6.

\section{The FPI transmission spectrum}

\begin{figure}
  \centering
  \includegraphics[width=0.45\textwidth]{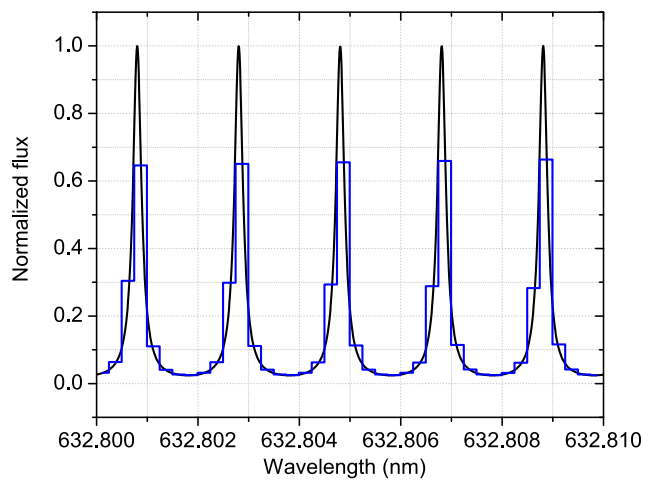}
  \caption{\label{fig:FPtrans}FPI interference pattern (black,
    $\mathcal{F}=10$) sampled by the astronomical spectrograph (blue)
    with a resolving power of R=80,000.}
\end{figure}

To reach the level of thermal and mechanical stability required for
sub-m/s accuracy is a challenging task in astronomical Doppler
experiments. In an echelle spectrograph, frequency calibration must be
provided by a spectral standard recorded on the very same detector as
the one optimized for astronomical observations. Therefore, spectral
resolving power and light intensity cannot be chosen arbitrarily.  For
the purpose of frequency calibration in astronomy, FPIs have several
advantages over other calibration sources, their main disadvantage
being that they do not inherently provide an absolute frequency
standard.

\begin{figure*}
  \centering
  \includegraphics[width=.9\textwidth,bbllx=0,bblly=0,bburx=830,bbury=514]{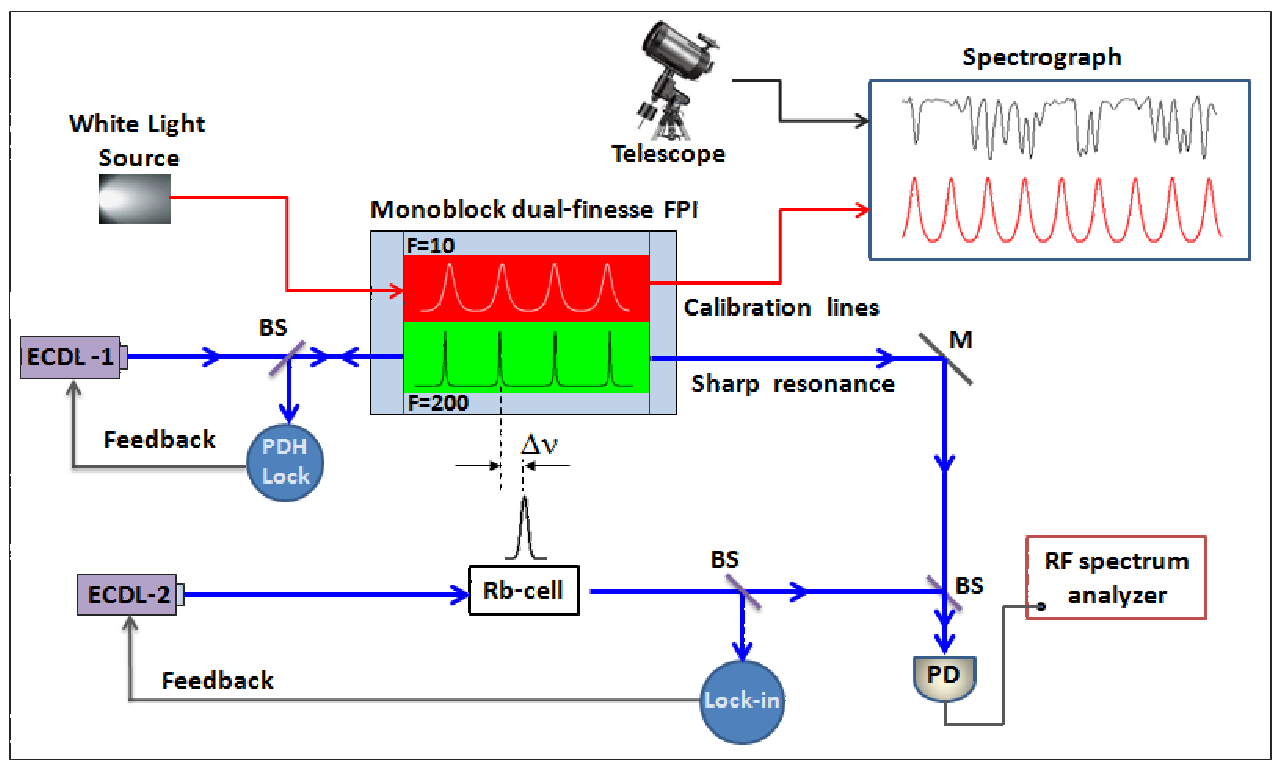}
  \caption{\label{fig:concept} Scheme of the proposed laser-lock system
    for tracking the FPI cavity. A white light continuum source illuminates the
    low-finesse Channel-1 interferometer (red block) to produce calibration
    lines for an Echelle spectrograph. The high-finesse Channel-2
    (green block) provides an external reference for locking the frequency of
    the diode laser ECDL-1 to the FPI cavity.  The RF-beat signal produced from
    interference between the Rb-locked ECDL-2 and the cavity locked ECDL-1 is
    recorded by high-speed photodetector PD. The cavity drift is to be
    inferred from the change in beat frequency.}
\end{figure*}

When illuminated with a white light source the gain-free FPI produces
a series of discrete cavity resonance lines separated by a fixed
frequency $\mathrm{\Delta \nu}_{\mathrm{FSR}}=\mathrm{c}/(2nl)$, where
$n$ is the refractive index of the cavity medium and $l$ the length. A
model spectrum of an FPI is shown in Fig.\,\ref{fig:FPtrans}. The main
specifications for an optimal FPI frequency calibration spectrum are
defined by the characteristics of astronomical spectrographs, which
are the following:
\begin{itemize}
\item The FPI lines should span the full wavelength range of the
  spectrograph and allow as much light as possible to enter
    the spectrograph. This entails the need for high-reflectance
  broadband coated cavity mirrors and a trade-off between
    peak width and the amount of light available for
    calibration. These requirements limit the reflectivity to roughly
  $R\simeq70\%$, resulting in rather low values for the finesse
  $\mathcal{F}=\pi R^{1/2}/(1-R) \!\la\!10$.
\item The FWHM of the transmission peaks should be close to (or below)
  the spectral resolving power of the instrument. The FWHM is
  determined by the finesse $\mathcal{F}$ of the interferometer
  through FWHM~=~FSR$/\mathcal{F}$, i.e, the free spectral range
  divided by the finesse. With the numbers used above, a typical FWHM
  of such an FPI is smaller than the spectrograph resolution, but is
  still several hundred m\,s$^{-1}$ to one km\,s$^{-1}$.
\item For optimal sampling, the FSR, i.e., the spacing between
  individual lines should be on the order of $3\times$\,FWHM of the
  spectrograph resolution. This is a compromise between spectral
  resolving power, transmission peak contrast, and maximizing the
  gradient of the signal over a wide spectral range. Wider spacing
  results in imperfect sampling (spectral regions with no frequency
  information), while too narrowly spaced lines cannot be resolved and
  lead to diminished contrast.
\end{itemize}

In general, the total line-position error $\sigma_\nu$ from photon
noise in an FPI transmission spectrum with $N$ comb lines, whose FWHM
is sampled by $p$ pixels each, can be approximated by \citep{bra87}

\begin{equation}
  \label{eq:sigmanu}
  \sigma_\nu=\frac{A}{\sqrt{Np}}\frac{\mathrm{FWHM}}{\mathrm{S/N}}.
\end{equation}
Here, S/N is the peak signal-to-noise ratio and $A$\,$(<1)$ a
normalization constant that depends on the shape of the line
profile. According to Eq.\,(\ref{eq:sigmanu}), a spectrum of an FPI
with $N\approx\!10^4$ lines, $p=10$, S/N\,=\,100, and a FWHM of
1.5\,GHz (15\,GHz FSR and $\mathcal{F}=10$) would reduce the total
frequency offset uncertainty to approximately 3\,cm\,s$^{-1}$ at
$\lambda = 600$\,nm and 6\,cm\,s$^{-1}$ at $\lambda = 1200$\,nm. This
performance is easily achieved in currently existing spectrographs and
meets the required precision of almost all astronomical RV measurement
carried out today. It shows that FPI spectra in general are well
suited to the frequency calibration of astronomical RV experiments.

The operational success of the FPI method depends crucially on the
dimensional stability of the cavity. Any perturbations in FPI cavity
length $L$ or refractive index $n$ would contribute as a shift in the
transmission peaks. For example, to achieve an RV precision of
10\,cm\,s$^{-1}$ with an FPI that has a free spectral range (FSR) of
15~GHz, the cavity length must be stable to roughly
$10^{-10}$\,cm. For low-expansion material (thermal expansion
coefficient $\alpha\sim10^{-8}$\,K$^{-1}$), this translates to a
temperature stability of a few mK \citep{sch12}. To further minimize
the effect of a varying refractive index $n$ on the optical path
length, the FPI must be placed inside a mechanically and thermally
stable evacuated chamber. Residual drifts in cavity length can be
estimated indirectly from rms temperature fluctuations measured inside
the vacuum chamber. Such measurements also have errors arising from a)
the uncertainties associated with the thermal expansion coefficients
and b) the placement of the temperature sensors away from the optical
path of the cavity.

A fundamental requirement for long-term stability is that the
dispersion of the cavity is constant with time. The change in mirror
dispersion can result from wavelength-dependent instabilities of the
optical path or from changes in the optical properties of the
coating. The former is minimized by avoiding any color-dependent
optical elements. For the latter, mechanical stress (specific to
coating process), aging, and light-induced damage (mostly from high
power lasers) are known to alter the reflectance of the mirror over
time \citep{Ennos:66, den14}. With an FP housed in a thermally
controlled evacuated chamber, we believe dispersion variations would
not be a major concern within a few years of operation, but
experimental prove of this assumption is not available yet. All
together, it is an extremely challenging problem to design an FPI that
can fulfill the stability requirements for cm\,s$^{-1}$ accuracy over
several years, which are the typical timescales for astronomical
experiments. A solution to this problem is to accept that the FPI will
not be stable at the required level of precision and to accurately
determine the amount of the drift, i.e., to track the variation of the
mirror distance (but still assuming that the dispersion is constant).
We propose to solve this problem by exploiting the metrology precision
available from optical frequency standards.

\section{The laser-lock concept}

The requirements that FPI stability reaches cm\,s$^{-1}$ long-term
accuracy by a long way exceeds today's technological possibilities. An
alternative to perfect stability is to accurately track this drift. A
straightforward way of implementing this is to cross-calibrate the FPI
spectrum with the primary wavelength standards used in astronomy, for
example hollow-cathode lamps or gas absorption cells. This can be done
using the astronomical spectrograph, and this strategy could improve
most of today's standard wavelength calibration procedures. The main
improvements are the clean FPI spectrum useful for continuous drift
check measurements and interpolation of the standard wavelength
solution in areas where primary standards have no spectral
features. This strategy, however, is fairly indirect and causes a
large overhead in calibration exposures, and it relies on the use of
other primary standards with all their shortcomings and measured in
the astronomical spectrograph. More direct drift-tracking of the FPI
transmission spectrum without involving the astronomical spectrograph
would be highly desirable.

Spectroscopic measurements based on spatial interference (at the
grating) measure wavelengths rather than frequencies of the
electromagnetic waves.  The resolution of dispersion-based instruments
invariably suffers from unavoidable wavefront aberrations in the
optical system and from the vast difference between the accuracy
desired (fractions of wavelengths) and the physical dimensions of
optical components. Frequency-measuring techniques, on the other hand,
have resulted in accuracies that are several orders of magnitude
higher than any wavelength measurement carried out with spectrographs
or interferometers \citep[e.g.,][]{Udem02}. Advances in laser
metrology allow us to connect the precision of frequency standards to
other physical quantities, including the measurement of the
dimensional stability of instruments \citep{rie98}. It is this
advantage that is utilized to reach accuracies achieved, for example,
by the laser-comb technology.

Employing the advantage of frequency standards, we intend to use an
optical heterodyning technique to directly measure the FPI cavity
drift. We propose to track the center position of one FPI resonance
peak by comparing it to a standard frequency provided by an external
standard, such as the $^{87}$Rb~$D_2$ line. The proposed setup is
shown in Fig.\,\ref{fig:concept}. Light from the telescope is fed into
the astronomical spectrograph together with the light comb from the
FPI (upper part of Fig.\,\ref{fig:concept}). The two frequencies of
one FPI resonance peak and the frequency standard (Rb cell) are locked
to a frequency stabilized external cavity diode laser (ECDL) operating
at 780\,nm; at ECDL-1, the laser is locked to the cavity after being
shifted by an electro-optical modulator, at ECDL-2 it is locked to
Doppler-free hyper-transitions of the Rb atom.

The locking mechanism ensures that any systematic or random drift in
cavity length is transferred to the frequency change in the ECDL-1,
and it accurately ties the frequency of ECDL-2 to a well-defined
physical standard. The beat between the two laser frequencies is then
measured at the photodetector (PD). The temporal interference between
two overlapping coherent fields with slightly different optical
frequencies ($\nu_{1,2}\sim 10^{14}$~Hz) from ECDL-1 and ECDL-2
creates a radio-frequency (RF) beat signal at much lower frequency
($\Delta f = |\nu_{1}-\nu_{2}|\approx 10^{6}\!-\!10^{9}$~Hz). The
amplitude and phase of the beat signal is easily measured with a high
speed PD. The output of the PD is processed electronically by an RF
spectrum analyzer or a digital frequency counter. Since the atomic
transitions are largely insensitive to external factors, the frequency
of ECDL-2, once locked to the Rb-cell, remains stable. Any change in
the beat signal would result from a frequency shift in ECDL-1
following the FPI cavity drift.

\begin{figure}
  \centering
  \includegraphics[width=0.49\textwidth]{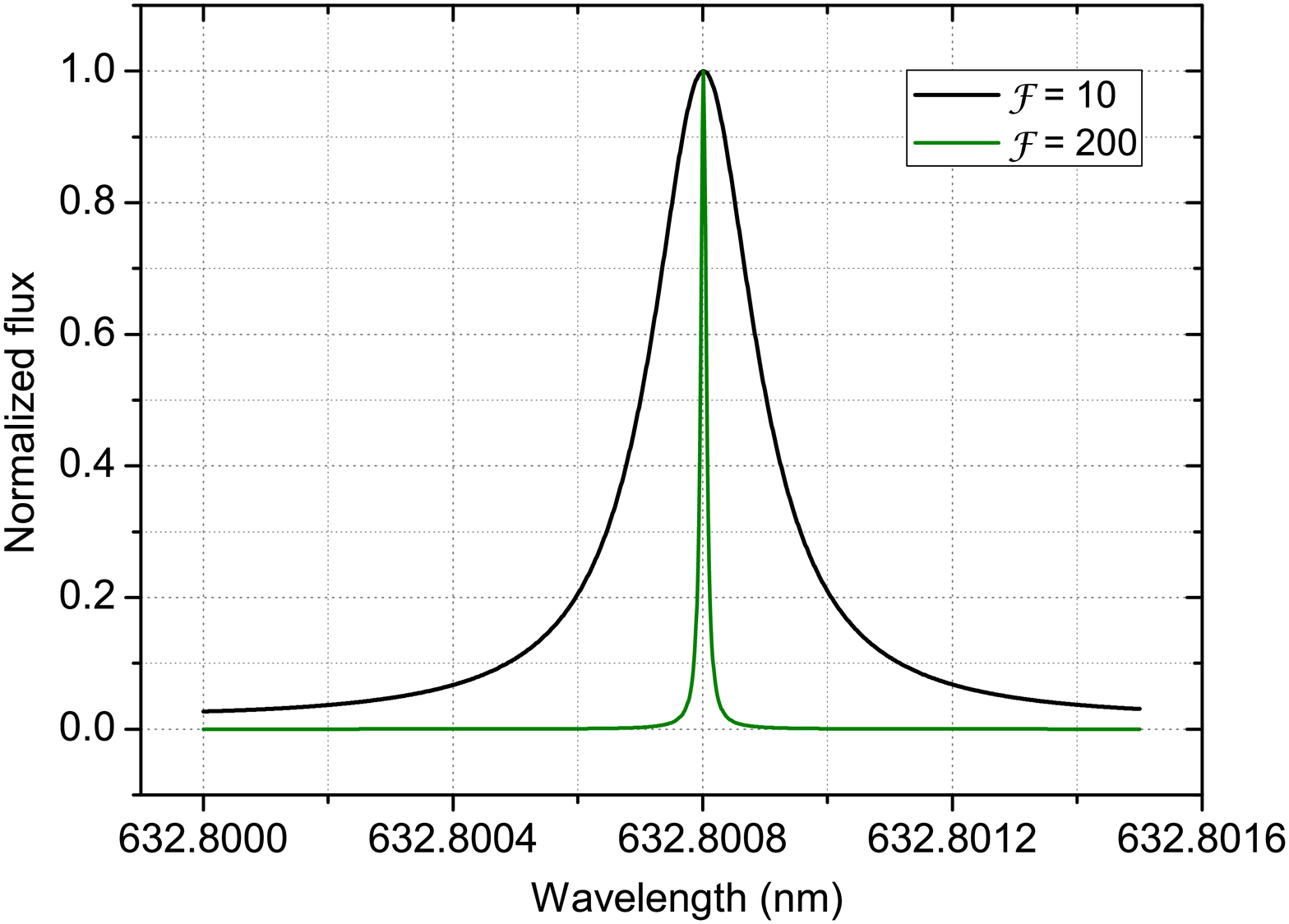}
  \caption{\label{fig:FPhighF}Comparison between a broad line (black
    curve, $\mathcal{F}=10$) from a low-finesse FPI and a narrow line
    (green curve, $\mathcal{F}=200$) from a high-finesse FPI.}
\end{figure}

The proposed concept foresees tracking of the FPI, which in principle
does not require very high stability. Nevertheless, in practice,
passive stabilization of the FPI (temperature and pressure) is needed
to ensure that the cavity does not drift more than a mode-hop free
scanning range of the diode laser, which is typically 5--6\,GHz.
Beyond this range, the laser frequency could suddenly jump to random
values, breaking the lock with the consequence of losing the
tracking. A larger cavity drift would also create an ambiguity in
keeping the cavity tied to Rb-line within 1/2 FSR.  It is technically
possible to increase the mode-hop free tuning range of the laser (e.g,
by using current feedforward), but a freely drifting cavity at ambient
temperature and pressure would place unrealistic demands on the speed
and bandwidth of photodetector system, thus increasing the complexity
and overall cost beyond any practical limits.

\section{The Dual-finesse FPI cavity}

A dual-finesse FPI cavity is similar to regular FPI, but it has a
coating with wavelength-dependent reflectivity. For example, a
reflectivity of 99\,\% can be reached around 780\,nm, the wavelength
range useful for standard Rb cells, and falls off to 70\,\% over wide
wavelength ranges. This cavity has low finesse ($\mathcal{F} \approx
10$) everywhere but around 780\,nm where its transmission peaks will
be very sharp ($\mathcal{F} = 200$). As noted earlier, the
high-finesse mirrors usually have narrow wavelength ranges and are
therefore not suitable for astronomical use where the entire spectral
range should be calibrated. On the other hand, the accuracy with which
the laser beam can be locked to the FPI cavity is a function of the
transmission peaks' FWHM. This accuracy is substantially lower than
the overall zero-point precision of the FPI because the laser beam is
locked to one single peak alone but not to the entire comb ($N = 1$ in
Eq.\,\ref{eq:sigmanu}). An FPI with a frequency-selective finesse is a
useful combination of coverage and precision. It has the great
advantage that the light used for radial velocity calibration and the
light for frequency locking follows an identical optical path that
eliminates uncertainties possibly introduced by solutions using
physically different surfaces.

Figure\,\ref{fig:FPhighF} shows a comparison between a resonance line
from a $\mathcal{F}=10$ FPI and a $\mathcal{F}=200$ FPI. The position
of the second line can be measured at substantially higher accuracy,
which can be seen from Eq.\,\ref{eq:sigmanu}: $N = 1$ (i.e., only one
transmission peak is used), the transmission peak S/N is approximately
the same for different values of the cavity finesse, $\mathcal{F}$,
the number of pixels sampled per FWHM, $p$, and FWHM scale with FWHM,
i.e., $\sigma_{\nu} \propto \sqrt{\rm{FWHM}}$. As an example, we
assume that the locking is carried out at roughly S/N~$\approx
1000$. For an FPI with FSR~$ = 15$\,GHz this results in a precision of
roughly 1.5\,cm\,s$^{-1}$ for a $\mathcal{F} = 200$ FPI
(FWHM~=~75\,MHz) and 30\,cm\,s$^{-1}$ for a $\mathcal{F} = 10$ FPI
(FWHM~=~1.5\,GHz) at 600\,nm. The same accuracy is reached using an
FPI with an FSR of 7.5~GHz operated at 1200~nm.  The best S/N can be
achieved with a high {\it spectral power density} source.  A
single-line laser diode with 30~mW output is sufficient to reach this
high S/N within a relatively short time ($\sim$ minutes).

In our dual-finesse cavity, the FSR and linewidth in the low-finesse
range comb lines are optimized for use in astronomical spectra.  The
second part has an FSR set by this choice of parameters, but its
finesse is higher, so the FWHM is smaller to allow a more precise
determination of the transmission peak frequency.

For the success of the locking concept, it is crucial that the light
entering the FPI follows a well-defined and stable optical path. This
is best achieved with single-mode fibers (SMF), and for the FP
calibrator, the choice of SMFs probably is absolute necessity. For
effective mode matching, the laser beam and white light source must
couple completely into the fundamental spatial mode of the cavity
\citep{and84}. Any departure from this condition may produce
undesirable artifacts in the calibration lines and render the laser
locking scheme ineffective. The calibration lines, e.g., produced in
an FP cavity illuminated with multimode fiber (core diameter $>
10\,\mu$m) can be highly asymmetric \citep{sch14}. This asymmetry is
caused by small angular and/or transverse offset between the fiber
axis and the cavity axis and it is found to increase with the fiber
core diameter, producing RV errors in excess of several
100\,m\,s$^{-1}$.  For astronomical use (telescope $\rightarrow$
spectrograph in Fig.\,\ref{fig:concept}), SMFs are avoided mainly
because of the light losses and difficulty coupling light efficiently
in SMFs, but this limitation does not apply to our FPI scheme, at
least not until the white light leaves the FPI on its way into the
astronomical spectrograph. The necessity of SMFs for the white light
path requires high intensity of the light source, for example from a
white laser.

\section{Laser frequency control}

In the proposed setup shown in Fig.\,\ref{fig:concept}, two single
frequency laser beams are required to track the FPI cavity drift.  A
frequency stabilized ECDL can be constructed using off-the-shelf
components \citep[e.g.,][]{ric95,arn98}. A generic frequency
stabilization scheme for an ECDL is shown in
Fig.\,\ref{fig:freqstab}. Prior to frequency locking, it is necessary
to enhance the passive stability of the diode to realize single-mode
operation.

\subsection{Passive frequency stabilization}

\begin{figure}
  \centering
  \includegraphics[width=0.49\textwidth]{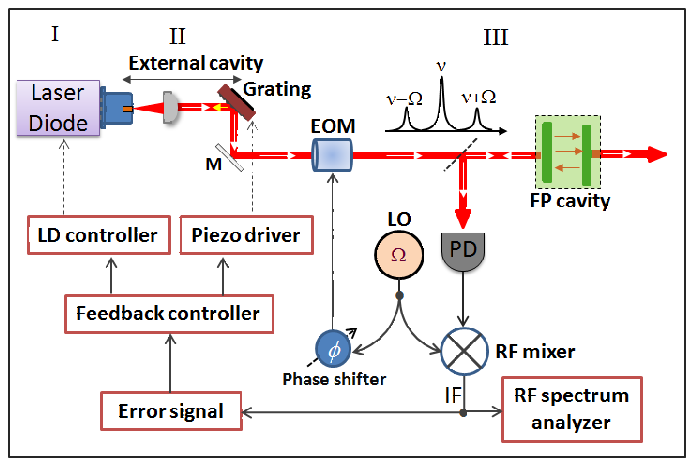}
  \caption{\label{fig:freqstab}Frequency stabilization of a
    semiconductor diode laser with the Pound-Drever-Hall technique
    \citep[see, e.g.,][]{dre83}. In Stage I, the linewidth of the
    laser is passively stabilized to $\sim$10\,MHz level by keeping
    the temperature and diode current stable. The grating feedback in
    Stage II reduces the linewidth to a few MHz. In Stage III, an
    active feedback mechanism reduces the linewidth further to about
    100\,kHz range by locking the laser frequency to the FPI cavity or
    the Rb-cell. Some other components, e.g., Faraday isolator and
    polarization optics, are not shown in the layout. LD: laser diode;
    EOM: electro-optic modulator; RF: radio frequency; PD:
    photodetector; LO: local oscillator.}
\end{figure}

The frequency of the laser diode is primarily determined by the gain
profile and longitudinal cavity modes that are normally separated by
100--200\,GHz. A typical index-guided laser diode operates at a cavity
mode that experiences the maximum gain. The laser frequency randomly
jumps from one mode to another owing to variations in cavity length
and the gain profile caused by random fluctuations in temperature and
injection current (phase noise). The frequency shift with temperature
and current is on the order of 100\,MHz/mK and $\sim3$\,MHz/$\mu$A,
respectively \citep{sal09-1}. By stabilizing the diode temperature to
a mK level and reducing the rms injection current noise to
$\leq1\,\mu$A, a single-frequency, mode-hop-free operation (Stage I in
Fig.\,\ref{fig:freqstab}) can be realized.

The phase-noise limit (also called Schawlow-Townes linewidth) of the
laser is proportional to the inverse square of the cavity length
\citep{hen86, osi87}. Therefore, the next step to minimize the phase
noise limit is to increase the cavity length $L$ of the laser. This is
achieved by adding an external reflector to the laser diode. Stage II
in Fig.\,\ref{fig:freqstab} shows the diffraction grating used in
Littrow configuration to form an extended cavity with the rear facet
of the diode. Many other variants of external cavities can be found in
the literature \citep[e.g.,][]{mro08}. The extended cavity effectively
reduces the phase noise, and the wavelength-selective feedback from
the grating narrows the laser linewidth further to a few 100\,kHz. A
continuous wavelength scan up to several GHz is easily achieved by
tuning the grating angle or changing the resonator length.

The linewidth of the ECDL is significantly narrower than the solitary
diode laser. However, owing to extreme temperature sensitivities and
reduced mode spacings, the laser output becomes highly prone to
mode-hopping. The laser itself is therefore not sufficient as absolute
wavelength standard, and further stability of the laser frequency can
only be ensured by actively locking the laser to a fixed frequency
reference.

\subsection{Active frequency stabilization}

To measure the FPI drift below 10~cm\,s$^{-1}$ and to monitor the
long-term stability of the cavity, the laser frequency spread should
be reduced to a few 100\,kHz or below. A stable linewidth close to
100\,kHz can be achieved by locking the laser to an external reference
frequency.  Deviations from the reference frequency are measured in
real time and corrected with an appropriate feedback. An active
stabilization requires fast electronic feedback control and a
frequency discriminator for suppressing the drifts and high frequency
technical noise caused by acoustics and mechanical vibrations
\citep{fox03}.

Fundamentally, all locking methods rely on generating an {\it error
  signal} that is proportional to the frequency deviations of the
laser source from the fixed reference. A steep slope about the lock
point result in a larger error and improved noise sensitivity. Various
modulation techniques are used for generating the error signal
\citep{neu09}. The choice of modulation frequency depends on the
characteristic linewidth of the spectral signal obtained from the
reference source \citep{bjo80}. For example, to lock the laser to the
Doppler-free Rb line (typical FWHM $\sim$ 6-8\,MHz), low modulation
frequencies of the diode current up to a few 100\,kHz are preferred.
Keeping the modulation frequency within the range of a few 100\,kHz is
necessary to prevent sideband interference with many closely spaced
hyperfine and crossover transitions in the atomic lines.

A very effective and high frequency modulation can be employed in a
Pound-Drever-Hall locking technique, represented as Stage III in
Fig.\,\ref{fig:freqstab}. PDH is a method for frequency stabilization
\citep{dre83} that produces a robust lock that is immune to intrinsic
fluctuations in laser power. It has a wide frequency-capture range and
is capable of suppressing high-frequency technical noise well beyond
the cavity response time. A detailed quantitative description of the
PDH technique is given by \citet{bla01}. Here, we give a brief summary
of the technique along with the key requirements relevant to our needs.

\begin{figure}
  \centering
  \mbox{\hspace{7.5mm}\includegraphics[width=0.45\textwidth]{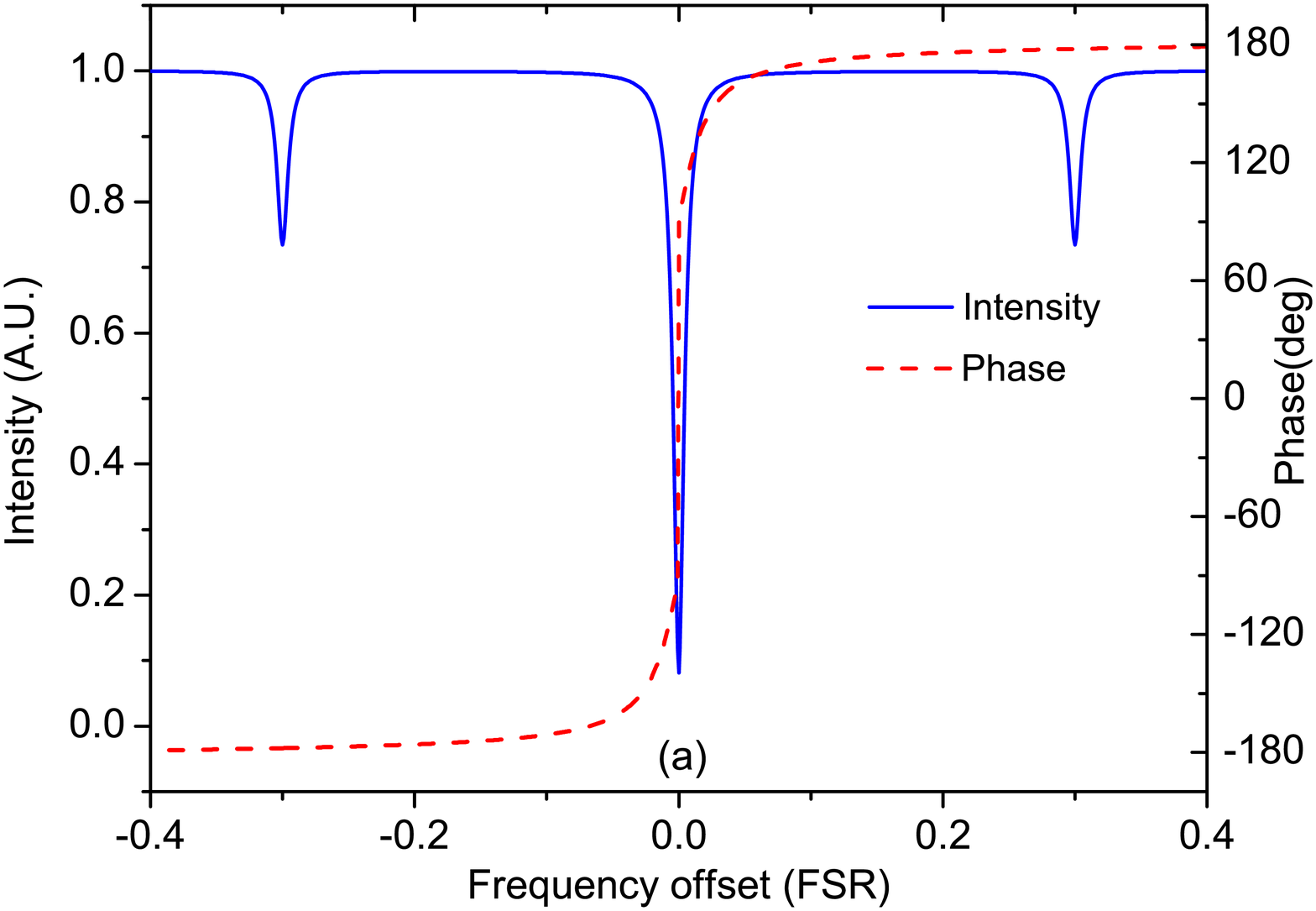}}\\
  \includegraphics[width=0.42\textwidth]{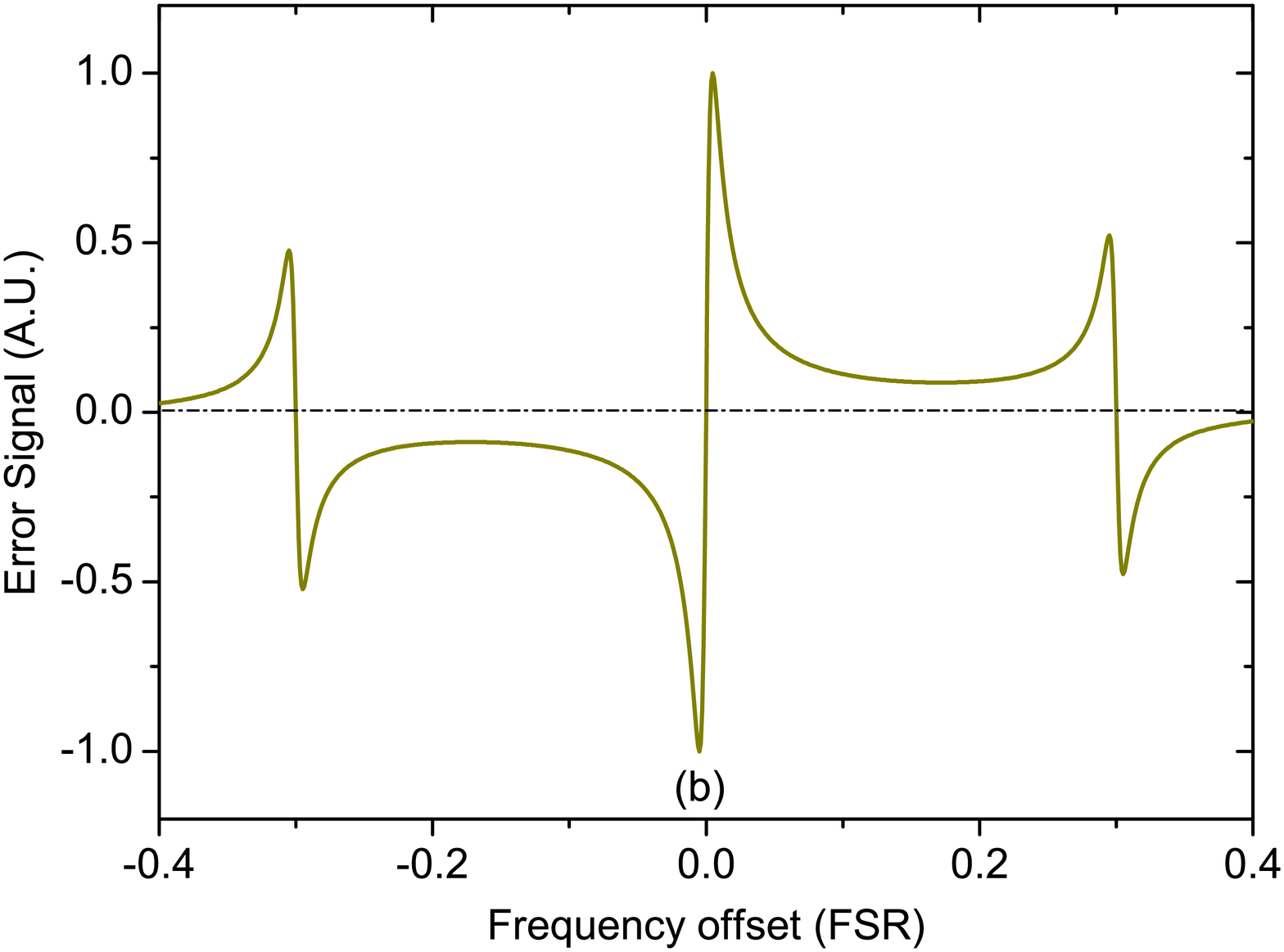}
  \caption{\label{fig:FPreflection}(a) Intensity (blue curve) and
    phase shift (red curve) of a reflected carrier from an optical
    cavity with $\mathcal{F}=200$. (b) The expected PDH error signal
    (the $\sin\mathrm{\Omega} t$-term, see text) with sidebands
    located within $30\%$ of the FSR. Laser is locked to the
    zero-crossing point of the error signal where the slope is
    steepest.}
\end{figure}

\begin{itemize}
\item The PDH lock operates on a light beam reflected from the FPI
  cavity. A monochromatic light beam whose frequency is an integral
  multiple of the cavity-free spectral range is completely
  transmitted, while an off-resonance beam is completely
  reflected. The reflected beam experiences a definitive phase shift
  that depends on its frequency with respect to the cavity
  resonance. The intensity of a reflected beam from an FPI cavity is
  plotted, together with its phase in Fig.\,\ref{fig:FPreflection}
  (upper panel). The phase of the reflected beam, shown only for the
  carrier frequency in Fig.\,\ref{fig:FPreflection}, has a sharp
  transition in the vicinity of the cavity resonance, which is used as
  a high-fidelity frequency discriminant to generate an
  error signal. For frequencies significantly \textit{above} and
  \textit{below} the cavity resonance, the phase shift, respectively,
  approaches a fixed value of +180$^\circ$ and -180$^\circ$.
\item To generate the error signal, the carrier frequency $\nu$ of the
  laser first is phase-modulated by a Pockels cell driven by a RF
  oscillator. The phase modulation of the laser beam creates two
  frequency sidebands at $\nu \pm \mathrm{\Omega}$. To ensure an
  extended lock range beyond the cavity linewidth, the modulation
  frequency $\mathrm{\Omega}$ must be a few times the FWHM of the
  cavity but should not exceed its FSR. For example, for a cavity with
  FSR~=~15\,GHz and FWHM~=~75\,MHz, sidebands at $\mathrm{\Omega} \sim
  \pm500$\,MHz would effectively overcome any technical noise in the
  laser frequency.
\item When a frequency-modulated beam is incident on the cavity, the
  two sidebands (that are outside the cavity resonance) are completely
  reflected with a $+180^{\circ}$ phase shift (the $\nu+\mathrm{\Omega}$
  component) and $-180^{\circ}$ (the $\nu-\mathrm{\Omega}$
  component). The carrier is also partly reflected if its frequency
  $\nu$ is slightly off resonance. Superposition of the reflected
  carrier and the sidebands generates a beat signal with
  $\sin\mathrm{\Omega} t$ and $\cos2\mathrm{\Omega} t$ terms at the
  photodetector \citep[see][for an exact derivation]{bla01}. The
  $\mathrm{\Omega}$ term represents the interference between carrier
  and sidebands, while the $2\mathrm{\Omega}$ term results from
  interference between the two sidebands. The amplitude and the sign
  of the $\mathrm{\Omega}$ term is extracted by RF-mixing of the
  photodetector output with the local oscillator signal (see
  Fig.\,\ref{fig:freqstab}).  The sign of the extracted signal
  determines which side of the resonance the laser frequency has
  drifted, while the amplitude provides the error signal proportional
  to the frequency deviation. A typical error signal is shown in the
  lower panel of Fig.\,\ref{fig:FPreflection} for the laser frequency
  excursions in units of cavity FSR.
\item The task of the feedback system is to use this error signal for
  adjusting the laser frequency by controlling one or more tuning
  parameters (injection current, temperature, or piezo-controlled
  cavity length) of the laser. In practice, a high servo gain and
  large bandwidth of the frequency tuning elements are needed to
  suppress the laser's intrinsic noise \citep{fox03}.
\end{itemize}

\subsection{Noise and performance limitations}

The accuracy of cavity drift measurements from the optical-beat signal depends
on the noise in the error signal.  At optimum locking condition (highest S/N),
the shot-noise limit determines the intrinsic frequency fluctuations of the
laser. These fluctuations cannot be distinguished from actual frequency
changes caused by the cavity drift. Laser frequency uncertainty associated with shot noise
can be expressed as \citep{dre83}
\begin{equation}
  \label{eq:FPunc}
  \delta\nu(\tau)\geq \mathrm{FWHM} \left(\frac{hc}{P_{0}T\eta\lambda}\right)^{1/2}\left(\frac{1}{\tau}\right)^{1/2},
\end{equation}
where FWHM is the cavity width, $P_{0}$ the laser power, $T$ is the
cavity transmission efficiency, $\eta$ the quantum efficiency of the
photodetector, and $(2\pi\tau)^{-1}$ the observation bandwidth. For an
FPI cavity of FWHM~=~100\,MHz, laser power $P_{0}=1$\,mW, $\lambda =
780$\,nm, $\eta = 0.9$, $T = 0.2$, and $(2\pi\tau)^{-1} = 1$\,kHz, we
calculate $\delta \nu(\tau) \approx 300$\,Hz as the frequency
stability limit from Eq.\,\ref{eq:FPunc}. This corresponds to a
precision of $\delta L / L = \delta \nu/\nu \approx 7.8 \times
10^{-13}$; v $\approx 0.2$\,mm\,s$^{-1}$. This shows that our proposed
locking scheme can determine the position of the FPI transmission
peaks with a precision that is safely beyond the calibration
requirements for all relevant Doppler experiments planned for the near
future.

The fundamental anchor of the calibration scheme is the stability of
the Rb atom. We are not aware of any specific evaluation of the
stability of a Rb cell over an extended period of time. Some attempts
have been made to identify and evaluate the impact of various
parameters and experimental conditions that can influence the
long-term, $>10^{5}$\,s, and short-term stability of Rb standards
\citep{ye96, van07,aff04}. The most notable factors that can perturb
the sub-Doppler lines of Rb atoms are fluctuations in the cell
temperature and a surrounding magnetic field.  Minimizing these
sources of noise is not a fundamental problem. The magnetic sub-level
shifting/splittings are a few MHz/G \citep{ste}, and they can be
largely eliminated by enclosing the Rb-cell inside a field-proof
casing capable of attenuating the effect of external magnetic field to
the level of a few $\mu$G. Furthermore, the hyperfine levels of the
$^{87}$Rb isotope are less susceptible to magnetic interference than
$^{85}$Rb. This advantage comes from the smaller nuclear spin ($I =
3/2$) of $^{87}$Rb resulting in fewer Zeeman sublevels as compared to
$^{85}$Rb85 ($I=5/2$). Finally, requirements on the thermal control of
the Rb cell are moderate because the temperature coefficient for line
shifts is about $10^{-11}$/K \citep{aff04}. A modest thermal stability
of $\sim 100$\,mK of the cell is sufficient to keep the
temperature-dependent drifts at least one order of magnitude below the
required RV precision.

\section{From precision to accuracy}

The locking scheme introduced above tracks variations of the FPI
spacer length precisely so that its uncontrolled drift can be
corrected on a cm\,s$^{-1}$ level. To compare measurements from
different instruments and to provide an absolute wavelength scale for
the astronomical measurements, it would be desirable to know the
\emph{absolute} frequency of the FPI transmission peaks. For this, we
need to identify the order of interference, $m$, of the peaks, which
is analogous to accurately determining the length of the spacer. In
this section, we investigate the question of whether an absolute
frequency scale can be established from an FPI spectrum in an
astronomical spectrograph \citep[see][]{MEISSNER:41,
  1999prop.book.....B}. The derivation of an absolute frequency scale
also works without the locking part but with reduced precision.

The transmittance, $T$, of an FPI can be described by
\emph{Airy's} formula,
\begin{equation}
  T = \frac{1}{1 + F \sin^2{\delta/2}},
\end{equation}
with $F = 4\mathcal{R} / (1 - \mathcal{R}^2)$ and $\mathcal{R}$ the
reflectivity. The phase, $\delta$, determines the positions of the
maxima according to the conditions of interference:
\begin{equation}
  \delta_{m} = \frac{4\pi}{\lambda_{m}} n l \cos{\theta} = 2 \pi m,
\end{equation}
with $\lambda_{m}$ the wavelength, $n$ the index of refraction, $l$
the length of the spacer, and $\theta$ the angle between the incoming
light and the interferometer surface. In the special case of $n=1$ and
perpendicular illumination ($\cos{\theta = 1}$), the condition for the
transmission peaks simplifies to
\begin{equation}
  \label{eq:lambdam}
  \lambda_{m} = 2 \frac{l}{m}.
\end{equation}

The transmission peaks of an FPI are equidistant in frequency space
($\nu$), and the spacing is
\begin{equation}
  \label{eq:spacing}
  \Delta \nu = \mathrm{FSR} = \frac{c}{2l},
\end{equation}
with $c$ the speed of light. Equation\,\ref{eq:spacing} shows the
immediate correspondence between spacer length $l$ and the peak
spacing, $\Delta \nu$~=~FSR. This means that if we can accurately
measure the distance between the transmission peaks, we can calculate
$l$ and find the absolute wavelength solution for our FPI spectrum.

We can uniquely identify the order number $m$ of a transmission peak
in the FPI spectrum if we can estimate $\lambda_{m}$ from
Eq.\,\ref{eq:lambdam} with an uncertainty $\delta \lambda_{m}$ that is
smaller than the distance between two peaks (FSR). We show in the
following that for this exercise, it is not necessary to determine the
position of two transmission peaks (defining FSR) through ultra-high
precision measurements, but that the accuracy provided with standard
calibration techniques in the astronomical spectrograph is
sufficient. An approximate wavelength scale can, for example, be
established from a hollow cathode lamp or a pen ray lamp in order to
define the position of at least two FPI transmission peaks. An
accuracy on the order of 100\,m\,s$^{-1}$ ($\sim$150\,MHz) is easily
achieved for the individual lines. The uncertainty in the difference
between the two peaks is then approximately $\delta \nu =
210$\,MHz. In relation to the FSR, this translates into an
unacceptable uncertainty of $\delta \lambda_{m} = 4000$\,km\,s$^{-1}$
for the absolute peak positions. However, the large lever arm of
$\Delta m \sim 10^4$ transmission peaks observed in an FPI spectrum
tailored for astronomical experiments solves this problem, because we
can select one transmission peak at the blue end of the spectrum and a
second at the red end, measure their positions, and count the number
of transmission peaks, $\Delta m$, between them (method of exact
fractions). The relative uncertainty $\delta l$ in the measurement of
the FPI spacing, $l$, can now be derived with the significantly
reduced relative uncertainty of
\begin{equation}
  \label{eq:FSRunc}
  \frac{\delta l}{l} = \frac{\delta \mathrm{FSR}}{\mathrm{FSR}} \frac{1}{\Delta m}.
\end{equation}
This is approximately $1.3 \times 10^{-6}$ for the numbers above,
which means an uncertainty in the \emph{absolute} frequency of $\delta
\lambda_{m} \approx 400$\,m\,s$^{-1}$. This is much lower than the
approximate 9\,km\,s$^{-1}$ spacing of FPI peaks. We can therefore
uniquely identify the mode of interference, $m$, for \emph{all}
transmission peaks if we know the exact position of one single peak
(using a locking scheme) and can obtain a wavelength solution with
relaxed accuracy for the rest of the wavelength range.

The identification of $m$, however, does not accurately define the
absolute wavelength of all peaks. Although our locking scheme provides
$l$ for the wavelength used for the locking scheme, we cannot assume
that $l$ is identical for all other wavelengths. The problem is that
the coating of our FPI has a very substantial extension relative to
$l$ and that phase changes can become important in interferometers
with dielectric coatings \citep[e.g.,][]{STANLEY:64, BENNETT:64,
  Lichten:85, Lichten:86}. In other words, $\delta l$ is not well
known for all peaks because the penetration depth of the light into
the coating (or the group velocity dispersion) is unknown. The
thickness of a coating is typically a few $\mu$m, which means that the
uncertainty in $l$, hence the absolute uncertainty of a transmission
peak, is up to $d/l \approx 10^{-6}/10^{-2} = 10^{-4}$ \citep[but its
effect on the FSR may be reduced using chirped
systems,][]{Szipocs:94}. The absolute uncertainty of individual
transmission peaks can therefore be as large as many km\,s$^{-1}$. On
the other hand, the penetration depth varies smoothly as a function of
frequency, which means that the FSR varies slowly, and the distance
between individual peaks varies slowly, too, which can help construct
the absolute wavelength scale. Nevertheless, the wavelength dependence
of FPI thickness and phase change upon reflection are serious problems
if FPIs are to be used for absolute wavelength calibration, and the
problem is even greater if the etalon has smaller spacing (for
example, fiber etalons). This problem has been discussed thoroughly in
the literature and finally led to the development of the LFC in which
every single comb peak is accurately defined.

\section{Summary}

\begin{table}
  \centering
  \caption{\label{tab:summary}Summary of typical parameters for FPI frequency calibration;
for the FPI we assume FSR = 15\,GHz, $\lambda = 600$\,nm}
  \begin{tabular}{lcccr}
    \hline
    \hline
    \noalign{\smallskip}
    Type of operation & $\mathcal{F}$ & $N$ & S/N & $\delta v$\qquad\ \\
    \noalign{\smallskip}
    \hline
    \noalign{\smallskip}
    ``Astro''                       &  10 & 10\,000 &  100 &  2.8\,cm\,s$^{-1}$\\
    single-peak ``Astro''           &  10 &    1    & 1000 & 28.5\,cm\,s$^{-1}$\\
    single-peak high- $\mathcal{F}$ & 200 &    1    & 1000 &  1.4\,cm\,s$^{-1}$\\
    \noalign{\smallskip}
    \hline
    \noalign{\smallskip}
  \end{tabular}
\end{table}

A Fabry-P\'erot interferometer operated with a white light lamp can
provide a large number of well-defined and uniform transmission peaks
that are suitable for frequency calibration in astronomical
spectrographs. We discussed advantages and challenges of FPI frequency
calibration and a laser-lock concept to precisely track frequency
offsets. A summary of the possible operation modes and precisions
reached in the measurement of frequency offsets is given in
Table\,\ref{tab:summary}.

The high number of available lines from an FPI tailored to
astronomical research can provide cm\,s$^{-1}$ precision if the FPI is
operated at a peak S/N around 100, which is a realistic value for
simultaneous wavelength calibration. The problem with such an FPI is
that stabilization of the cavity length to this level of precision is
impractical so that the cavity drift contributes a larger
uncertainty than the actual determination of the transmission peak
position in the astronomical spectrograph.

The solution we proposed is to track the cavity drift by comparing the
position of one transmission peak to an external standard, e.g., an
atomic Rb transition. The measurement of a single line peak position
in the same FPI that is used for the astronomical measurement cannot
be measured with m\,s$^{-1}$ precision at a S/N of 100, and the same
is true for any other spectral feature from external sources in the
spectrograph itself. This measurement has to be carried out externally
where it can be done at much higher S/N. For example, at S/N\,=\,1000
the uncertainty in the peak position is roughly 30\,cm\,s$^{-1}$,
which may already be useful for astronomical purposes. An extension to
this concept is to use a second FPI with higher finesse; our proposal
foresees a dual-finesse cavity with an optimized reflectivity that
varies with wavelength. With a finesse of 200 and a S/N of 1000 a
simple tracking mechanism can track the offset at a precision of
roughly 1\,cm\,s$^{-1}$ using a single line in a wavelength region
where the cavity has high reflectivity. A notch filter can be used to
prevent the high intensity of the laser from entering the
spectrograph. Another improvement is the implementation of a locking
mechanism, such as PDH. This ensures a measurement precision of the
drift between the FPI and the external standard that is only a
fraction of a mm\,s$^{-1}$.

Several other solutions that track the drift of the interferometer are
possible. For example, as a compromise between rather low finesse for
the broad (astronomical) range and relatively high finesse for
tracking, one may choose $\mathcal{F} \approx 50$ for the entire
wavelength range. Such an FPI can cover a wide wavelength range and
still provide relatively high single-line precision when measured at
high S/N. With a typical exposure time of several minutes, integration
times for the drift tracking are long enough to measure the drift at
much higher S/N than 1000, which relaxes the requirements for all
other parameters.

The wide wavelength range of an astronomical FPI also allows the mode
of interference to be determined for all transmission peaks. This is
possible when two transmission peaks can be measured that are
separated by a large difference in interference mode number; the
relative uncertainty of the FSR is effectively reduced by the large
lever arm from the high number of modes generated by the low-finesse
FPI. Thus, the absolute length of the spacer is measured at one
wavelength and the remaining uncertainty of all other peaks is reduced
to the uncertainty in the variation in spacer length as a function of
wavelength. This uncertainty, however, can be very substantial for
wavelength ranges far away from the locked mode. A potential solution
for astronomical purpose is to calibrate the FPI spectrum using an LFC
in a laboratory. This would define the length of the spacer at every
wavelength because the frequency of every peak can be measured and $m$
is known (although the accuracy of this calibration will be much lower
than the one reached in the locked mode). This information can be used
to determine the position of \emph{all} peaks from tracking one
transmission by a locking scheme. Ideally, this strategy would allow
absolute calibration using all transmission peaks of the FPI with an
uncertainty limited by the ability to cross-reference the FPI spectrum
to an external calibrator. This could be a cost-efficient way to apply
the absolute wavelength accuracy of an LFC to astronomical
spectrographs.

An externally tracked FPI can provide a light source for frequency
calibration that allows cm\,s$^{-1}$ precision measurements in
astronomical Doppler experiments.  For high-precision Doppler
measurements, the externally tracked FPI is a huge improvement over
traditional techniques like HCL calibration. The total cost of such a
system can be estimated to be much lower than 20\,\% of the typical
cost for an astronomical LFC. The FPI calibrator is a high-precision
calibration source that can become affordable for a wide range of
observatories. All technology is available so that the proposed
strategy is particularly interesting for upcoming experiments, such as
CARMENES \citep{2012SPIE.8446E..0RQ}, HZPF
\citep{2012SPIE.8446E..1SM}, and SPIRou \citep{2011ASPC..448..771A}.

\begin{acknowledgements}
  We thank the anonymous referee for a very helpful report. This research was
  supported by the European Research Council under the FP7 Starting Grant
  agreement number 279347. AR acknowledges research funding from DFG grant RE
  1664/9-1.
\end{acknowledgements}

\bibliographystyle{aa}
\bibliography{mybib}

\end{document}